\documentclass[prb,twocolumn]{revtex4-1}
\usepackage[table, dvipsnames]{xcolor}
\usepackage{graphicx}
\usepackage{amssymb}
\usepackage{dcolumn}
\usepackage{float}
\usepackage{bm}
\usepackage{multirow}
\usepackage{booktabs}
\usepackage[T1]{fontenc}
\usepackage[utf8]{inputenc}
\usepackage{lmodern}
\usepackage{color}
\usepackage{makecell}
\usepackage{tikz}
\usepackage{pgfplots}
\usepackage{mathtools}
\usepackage{physics}
\usepackage{amsmath}
\usepackage{setspace}
\usepackage{ragged2e}

\usepackage[colorlinks=true,linkcolor=red]{hyperref}

\newcommand{\mathleft}{\@fleqntrue\@mathmargin0pt}

\newcolumntype{M}[1]{>{\centering\arraybackslash}m{#1}}

\renewcommand{\i}{\mathrm{i}}

\DeclareMathAlphabet{\bi}{OML}{cmm}{b}{it}

\def\be{\begin{equation}}
	\def\ee{\end{equation}}
\def\bearr{\begin{eqnarray}}
	\def\eearr{\end{eqnarray}}

\begin{document}
	
	\title{Enhanced Gilbert Damping via Cubic Spin-Orbit Coupling at 2DHG/Ferromagnetic Insulator Interface}
	\bigskip
	\author{Sushmita Saha and Alestin Mawrie}
	\normalsize
	\affiliation{Department of Physics, Indian Institute of Technology Indore, Simrol, Indore-453552, India}
	\date{\today}

\begin{abstract}
	We investigate the enhancement of Gilbert damping at 2DHG/ferromagnetic insulator (FI) interfaces, where spin pumping from the FI layer injects spins into the 2DHG, and cubic Rashba spin-orbit coupling (RSOC) significantly boosts spin relaxation and spin-pumping efficiency compared to 2DEG systems. The dominant contribution to spin damping arises from interband transitions which does exhibits conductivity-like behavior as the temperature, \( T \to 0 \). Our results reveal that damping remains stronger than in 2DEG due to the persistent influence of cubic RSOC. The interplay between RSOC and magnon absorption broadens the spectral response, with the damping peak shifting more notably at higher temperatures. Stronger RSOC expands the magnon interaction phase space, thus widening the damping spectrum. A key observation emerges with the Fermi level (\(E_f\)): a finite \(E_f\) sustains spin imbalance and enhances damping, whereas \(E_f = 0\) suppresses it, unlike in 2DEG. The electric field tunability of RSOC enables real-time control over spin relaxation and angular momentum transfer, offering a pathway toward voltage-controlled spintronic devices. These findings highlight the superior potential of 2DHG for tailoring spin dynamics via electric and thermal effects.
\end{abstract}

	\email{amawrie@iiti.ac.in}
	\pacs{78.67.-n, 72.20.-i, 71.70.Ej}
	
	\maketitle
	
	\section{Introduction}

\begin{figure}[t]		\includegraphics[width=60.5mm,height=40.5mm]{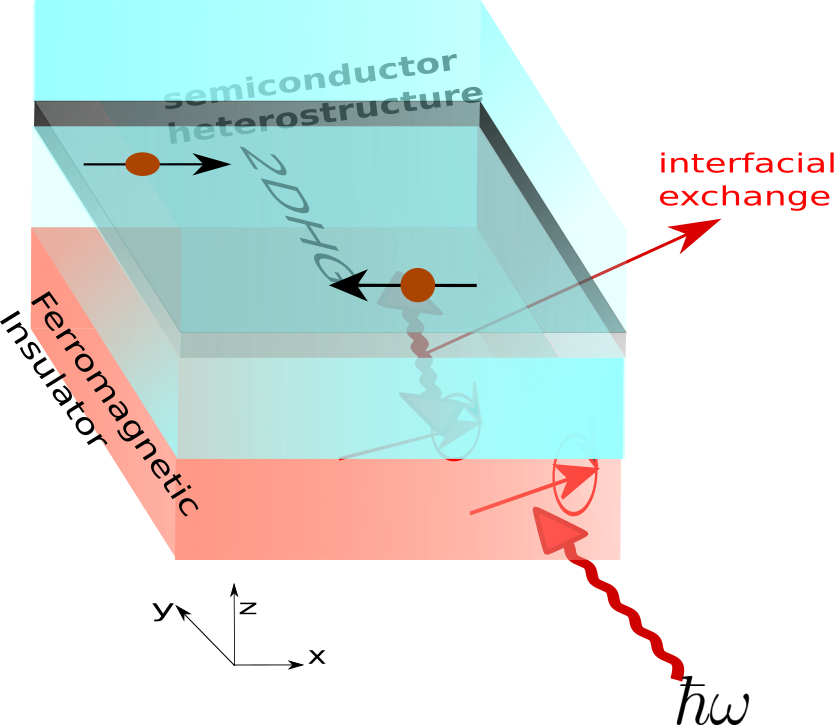}	
	\caption{Schematic of spin pumping from a ferromagnetic insulator layer into a 2DHG via interfacial exchange under microwave excitation.
	}
	\label{Fig1}
\end{figure}

The interaction between spin-orbit coupling (SOC) and magnetization dynamics at two-dimensional hole gas (2DHG)/ferromagnetic insulator (FI) interfaces (see Fig.[\ref{Fig1}]) presents a promising avenue for exploring spintronic phenomena. In particular, SOC in 2DHGs, formed at the interface of \textit{p}-type AlGaAs/GaAs heterostructures, plays a crucial role in influencing spin transport and magnetic interactions due to the \textit{p}-orbital nature of the charge carriers \cite{SOCex1,SOCex2}. The enhanced SOC in 2DHGs \cite{SOC1,SOC2} compared to two-dimensional electron gases (2DEGs) \cite{2DEG} leads to distinct magnetic phenomena when interfaced with an FI. The exchange interaction \cite{EI1,EI2,EI3} at this interface, involving spin-polarized electrons from the ferromagnet and SOC-coupled holes in the 2DHG, results in intricate magnetic behaviors that are crucial for various spintronic applications.

A key consequence of this exchange interaction is the enhancement of Gilbert damping, which governs energy dissipation in magnetization dynamics \cite{takahasi}. In 2DHG/FI heterostructures, strong cubic Rashba SOC \cite{2DHG1,2DHG2,2DHG3,2DHG4,2DHG5} facilitates efficient spin relaxation, further amplifying damping. Moreover, spin-pumping \cite{Qscattering}, where spin angular momentum is transferred from the precessing ferromagnet to the adjacent 2DHG, provides an additional mechanism for energy dissipation. At the interface, strong SOC enhances this spin transfer, leading to more efficient angular momentum dissipation. The combined effects of SOC-driven spin relaxation \cite{SOC spin pumping} and spin-pumping establish a highly efficient route for controlling magnetization dynamics, making these heterostructures promising for spintronic applications requiring fast and energy-efficient magnetic switching.

A series of seminal works have extensively explored various aspects of spin-pumping \cite{Kato1,Kato2,Kato3,Kato4,Kato5,Kato6,Kato7,Kato8}, significantly contributing to its theoretical modeling. Meanwhile, experimental studies \cite{expt1,expt2,expt3,expt4} have investigated spin-pumping effects in different material systems, providing deeper insights into the underlying physics.

In this work, we explore the enhancement of Gilbert damping at 2DHG/FI interfaces and demonstrate that it originates from strong cubic Rashba spin-orbit coupling (RSOC), which enables more efficient spin relaxation compared to conventional 2DEG systems. Our findings reveal that this enhancement is primarily driven by interband transitions which also exhibits conductivity-like behavior\cite{conductivity} at low temperatures. As the temperature rises, magnon contributions become significant, leading to spectral broadening and a temperature-dependent shift in the damping peak, ultimately affecting spin current\cite{spin current1,spin current2} through spin pumping.

Moreover, we identify the optical frequency range suitable for microwave excitation, ensuring efficient spin pumping and controlled damping enhancement. Our results also highlight the crucial role of the Fermi energy ($E_f$) in sustaining spin imbalance and enabling spin current injection. Notably, setting $E_f = 0$ in a 2DHG suppresses this effect, in contrast to 2DEG systems\cite{2DEG}. Additionally, we find that stronger RSOC enhances magnon interactions, broadening the spectral response by reducing magnon lifetime through increased scattering, thereby extending the damping over a wider frequency range.

Significantly, the ability to dynamically tune RSOC strength in 2DHGs via an external electric field allows real-time control over spin current dissipation. These findings position 2DHG-based heterostructures as promising platforms for voltage-controlled spintronics, opening new avenues for efficient spin manipulation and next-generation device applications.  

The structure of the paper is organized as follows: In Section [{\ref{sec1}}], we introduce the Hamiltonian that characterizes the composite system depicted in Fig. [{\ref{Fig1}}]. In Section [\ref{sec2}] we include a thorough derivation of the Gilbert damping factor, providing a detailed step-by-step calculation. Section [{\ref{sec3}}] is dedicated to presenting the results of the analysis, where we discuss the implications and significance of the findings. Finally, Section [{\ref{sec4}}] concludes the paper, offering a summary of the key insights and highlighting possible directions for future work.

\section{Total Hamiltonian for the Composite Structure in Fig. [\ref{Fig1}]}\label{sec1}
The total Hamiltonian (\( H_{\rm tot} \)) of the composite system (shown in Fig. [\ref{Fig1}]) under consideration comprises three key components, each representing a different aspect of the physical interactions at play in the system. These components are the Hamiltonian of the two-dimensional hole gas (\( H_{\rm{2DHG}} \)), the Hamiltonian of the FI (\( H_{\rm{FI}} \)), and the interfacial exchange Hamiltonian (\( H_{\rm{ex}} \)). 
\begin{eqnarray}
H_{\rm tot} ={H}_{\rm{2DHG}}+{H}_{\rm{FI}}+{H}_{\rm{ex}}
\end{eqnarray}
\subsection{Two-dimensional heavy hole gas}
The general form of the Hamiltonian, \( H_{\text{2DHG}} \), for a 2DHG with cubic spin-orbit interactions can be written in terms of the creation and annihilation operators for spin-up (\( c_{\textbf{k},\uparrow}^\dagger \;\&\; c_{\textbf{k},\uparrow}\)) and spin-down (\( c_{\textbf{k},\downarrow}^\dagger\;\&\; c_{\textbf{k},\downarrow} \)) holes in momentum space. The Hamiltonian for this system, including the cubic Rashba\cite{Rashba1,Rashba2} and Dresselhaus\cite{Dre1,Dre2} spin-orbit interactions, is:
\[
{H}_{\rm{2DHG}} = \sum_{\textbf{k}} \begin{pmatrix}
c_{\textbf{k},\uparrow}^\dagger & c_{\textbf{k},\downarrow}^\dagger
\end{pmatrix}
\mathcal{H}_{\text{2DHG}}({\bf k})
\begin{pmatrix}
c_{\textbf{k},\uparrow} \\
c_{\textbf{k},\downarrow}
\end{pmatrix}
\]
Here, \( \mathcal{H}_{\text{2DHG}}({\bf k}) \) is a \(2\times2\) matrix that describes the effective Hamiltonian for the spin-orbit coupled 2DHG at momentum \( \textbf{k} \), and it incorporates both the Rashba and Dresselhaus spin-orbit interactions as shown below
\begin{eqnarray}
\mathcal{H}_{\text{2DHG}}({\bf k})& =& \left[\zeta(\textbf{k})-E_f\right] \sigma_0 + \frac{\i \alpha}{2} ({k_-}^3 \sigma_- - {k_+}^3 \sigma_+) \nonumber\\
&-& \frac{\beta}{2} (k_- k_+ k_- \sigma_+ + k_+ k_- k_+ \sigma_-).
\end{eqnarray}

\noindent Here \( k_\pm = k_x \pm i k_y \) are the complex momentum components, and \( \sigma_+ / \sigma_- \) are the spin raising/lowering operators, respectively. The first term represents the kinetic energy of the holes, with \( \zeta(\textbf{k})={\hbar^2 k^2}/{2m} \) being the standard free-hole energy of mass, \(m\) and \(E_f\) the Fermi level. The second and third terms describe the Rashba and Dresselhaus spin-orbit interactions, respectively, where the coupling constants \( \alpha \) and \( \beta \) govern the strength of each interaction.
The cubic dependence of the spin-orbit coupling terms results from the \(p\)-orbital nature of holes in the valence band, leading to more complex spin dynamics compared to systems with linear spin-orbit coupling\cite{linear}. 
These interactions play a crucial role in governing spin relaxation processes, spin-momentum locking, and magnetization dynamics in 2DHG systems, directly impacting the efficiency of spin pumping from a FI into the 2DHG, which is the focus of this paper.
\begin{figure}[b]		\includegraphics[width=55.5mm,height=40.5mm]{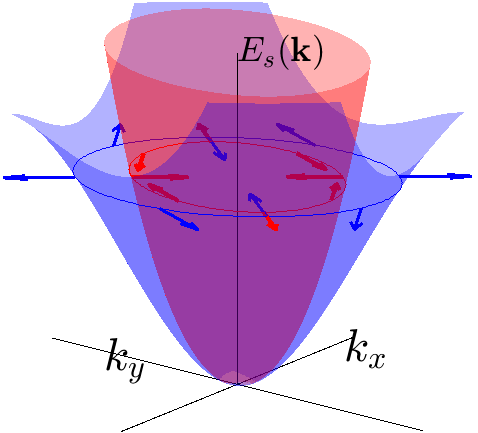}	
	\caption{The figure illustrates the energy dispersion $E_s(\mathbf{k})$ in the presence of spin-orbit coupling, where the red surface corresponds to $s = +1$ and the blue surface to $s = -1$. The arrows indicate the spin texture on the Fermi contours, showing opposite chiralities for the two branches.
	}
	\label{Fig2}
\end{figure}

The expression for the magnitude of the SOC field \( \textbf{h}_{\rm{SOC}}({\bf k}) \) is given as:
\begin{equation}
|\textbf{h}_{\rm{SOC}}({\bf k})| ={k}^3\Delta(\phi),
\end{equation}
where \(\Delta(\phi)=\sqrt{\alpha^2+\beta^2-2\alpha\beta\sin(2\phi)}\) with the angle \( \phi \) typically the azimuthal angle in momentum space. The form of this equation indicates that the strength of the spin-orbit coupling depends not only on the parameters \( \alpha \) and \( \beta \), but also on the angle \( \phi \), suggesting that the coupling has a directional dependence in momentum space.
The SOC field is described as a vector:
\begin{equation}
\textbf{h}_{\rm{SOC}}({\bf k}) =\begin{pmatrix} h_x \\ h_y \\ 0 \end{pmatrix}=\begin{pmatrix}
{k}^3 \, (-\alpha\sin3\phi + \beta\cos\phi)
\\
{k}^3 \, (\alpha\sin3\phi + \beta\cos\phi)\\
0
\end{pmatrix}.
\end{equation}
These components describe the projection of the SOC field in the \( x \)- and \( y \)-directions in momentum space, and their functional dependence on \( \phi \) suggests a complex, angle-dependent spin texture as illutrated in Fig.[\ref{Fig2}]. The terms involving \( \sin(3\phi) \) and \( \cos(\phi) \) reflect the interplay between the Rashba and Dresselhaus interactions, which have distinct symmetries in momentum space.
Finally, the energy of the system, $E_{s}(\textbf{k})$, is expressed as:
\begin{equation}
 E_{s}(\textbf{k}) = \zeta(\textbf{k}) + s \, |\textbf{h}_{\text{SOC}}({\bf k})|,
\end{equation}
where \( s \) indicates the spin polarization. The second term, \( s \, |\textbf{h}_{\text{SOC}}({\bf k})| \), with \(s=\pm\) describes the energy shift to the total energy due to the spin-orbit coupling.


We now introduce the retarded Green's function for a 2DHG through the following differential equation: 
\begin{equation}
[E - H_{\text{2DHG}}(\mathbf{r})] \, \mathcal{G}_0(\mathbf{r}, \mathbf{r}^\prime) = \delta(\mathbf{r} - \mathbf{r}^\prime),
\end{equation}
where $\mathcal{G}_0(\mathbf{r}, \mathbf{r}^\prime)$ represents the Green’s function for the pure system. This Green’s function is symmetric under the exchange of spatial coordinates, satisfying $\mathcal{G}_0(\mathbf{r}; \mathbf{r}^\prime) = \mathcal{G}_0(\mathbf{r}^\prime; \mathbf{r})$. The Hamiltonian of the system is modified to include impurity scattering, expressed as $H_{\rm{2DHG}}^\prime \Psi = \mathcal{E} \Psi$, where $H_{\rm{2DHG}}^\prime = H_{\rm{2DHG}} + \mathcal{V}_{\text{imp}}$. The impurity potential $\mathcal{V}_{\text{imp}}$ is modeled as 
\begin{equation}
\mathcal{V}_{\text{imp}} = u \sum_{i \in \text{imp}} \sum_{\sigma} \Psi^\dagger (\mathbf{R}_i) \Psi(\mathbf{R}_i),
\end{equation}
with the field operator $\Psi(\mathbf{\textbf{R}}_i)$ expressed in terms of momentum eigenstates as 
\begin{equation}
\Psi(\mathbf{R}_i) = \frac{1}{\sqrt{\Omega}} \sum_k c_{\textbf{k},\sigma} e^{i \mathbf{k} \cdot \mathbf{R}_i}.
\end{equation}
Here, $\mathbf{R}_i$ denotes the positions of the impurities and $\sigma$ represents the spin index, and \(\Omega\) the surface area of the 2DHG. The Green's function for the system including impurity scattering satisfies the equation 
\begin{equation}
[\mathcal{E} - H^\prime_{\text{2DHG}}] \, \mathcal{G}(\mathbf{r}, \mathbf{r}^\prime, \tau) = \delta(\mathbf{r} - \mathbf{r}^\prime, \tau).
\end{equation}

Starting with the definition of the retarded Green's function in real space and time, it can be expanded as 
\begin{equation}
\mathcal{G}(\mathbf{r}, \mathbf{r}^\prime, \tau) = \sum_{\mathbf{k}, \mathbf{k}^\prime} \langle \mathbf{r} | \mathbf{k} \rangle \mathcal{G}(\mathbf{k}, \mathbf{k}^\prime, \tau) \langle \mathbf{k}^\prime | \mathbf{r}^\prime \rangle.
\end{equation}
By performing a basis transformation, the Green's function in momentum space is expressed as  
\begin{equation}
\mathcal{G}_{\sigma \sigma^\prime}(\mathbf{k}, \tau) = -\hbar^{-1} \langle c_{\textbf{k},\sigma}(\tau) c_{\textbf{k},\sigma}^\dagger \rangle,
\end{equation}
where the thermal expectation value is computed using 
\begin{equation}
\langle c_{\textbf{k},\sigma}(\tau) c_{\textbf{k},\sigma}^\dagger \rangle = \frac{\text{Tr}[\rho c_{\textbf{k},\sigma}(\tau) c_{\textbf{k},\sigma}^\dagger]}{\text{Tr}[\rho]},
\end{equation}
with the density matrix defined as $\rho = e^{-\beta \mathcal{H}^\prime_{\text{2DHG}}}$. The time evolution of the operators is governed by 
\begin{equation}
c_{\textbf{k},\sigma}(\tau) = e^{H^\prime_{\text{2DHG}} \tau / \hbar} \, c_{\textbf{k},\sigma} \, e^{-H^\prime_{\text{2DHG}} \tau / \hbar},
\end{equation}
valid for $0 < \tau < \hbar \beta$.
The Green's function can be transformed from the time domain to the frequency domain via a Fourier transform, providing information about the possible energy states of the system. The resulting Green's function in the frequency domain is 
\begin{equation}
\mathcal{G}(\mathbf{k}, i\omega_n) = \hbar \beta \int_{0}^{\beta} d\tau \, e^{i\omega_n \tau} \mathcal{G}(\mathbf{k}, \tau).
\end{equation}
For the pure system, the Green's function in terms of Matsubara frequencies is given by 
\begin{equation}
\mathcal{G}_0(\mathbf{k}, i\omega_m) = \frac{(i \hbar \omega_m - \zeta(\textbf{k})) \boldsymbol{\sigma}_0 - \mathbf{h}_\text{SOC} \cdot \boldsymbol{\sigma}}{(i \hbar \omega_m - E_+(\textbf{k}) )(i \hbar \omega_m - E_-(\textbf{k}) )},
\end{equation}
where $i\omega_m = (2n+1)\pi / (\hbar \beta)$ are the Matsubara frequencies for fermions. 
In real system, the presence of impurities modifies the Green's function due to scattering effects. Incorporating these effects, the Green's function becomes \begin{equation}
\mathcal{G}(\mathbf{k}, i\omega_m) = \frac{\big[i \hbar \omega_m - \zeta(\textbf{k}) + i\Gamma/2 \, \mathrm{sgn}(\omega_m)\big] \boldsymbol{\sigma}_0 - \mathbf{h}_\text{SOC} \cdot \boldsymbol{\sigma}}{\prod_{s=\pm} \big[i \hbar \omega_m - E_s(\textbf{k}) + i\Gamma/2 \, \mathrm{sgn}(\omega_m) \big]},
\end{equation}
where $\Gamma = 2\pi n_{\text{imp}} u^2 \mathcal{D}(E_f)$ represents the level broadening due to impurity scattering. The detailed derivation for the above equation can be found in the Appendix [\ref{AppI}]. This formulation highlights the interplay of impurity scattering\cite{Lee} and spin-orbit coupling on the electronic properties of the system.


\subsection{Ferromagnetic Insulator}
A ferromagnetic insulator is a material where the magnetic moments of the constituent atoms or ions align parallel to each other due to strong exchange interactions, resulting in a net magnetization. In such systems, magnons\cite{magnon} i.e. quantised spin waves serve as the primary carriers of spin angular momentum, facilitating the spin transport without charge transport. Unlike metals, where the free electron spins contribute to ferromagnetism, in insulators, the magnetic moments are usually due to localized spins of the ions or atoms. 
The spin at each lattice site \(l\) can be represented by a vector \(\mathbf{S}_l\). In a simplified model, the average of this spin vector can be expressed in terms of spherical coordinates:
\begin{eqnarray}
\langle\mathbf{S}_l\rangle = (\langle\mathbf{S}_x\rangle^l, \langle\mathbf{S}_y\rangle^l, \langle\mathbf{S}_z\rangle^l)^\prime = S_0 \textbf{m},
\end{eqnarray}
where \(\textbf{m}=\left( \cos \theta,\sin \theta, 0\right)\).
Here, \(S_0\) is the magnitude of the spin, and \(\theta\) is the angle it makes with a reference axis. To facilitate calculations involving rotated coordinate systems, we introduce a rotation matrix \(\mathcal{R}(\theta, \phi)\) that relates the original spin components to the new ones \((S_{x^\prime}, S_{y^\prime}, S_{z^\prime})\):
\begin{eqnarray}
\begin{pmatrix}
S_{x^\prime}\\
S_{y^\prime} \\
S_{z^\prime}
\end{pmatrix}
=
\mathcal{R}(\theta, \phi)
\begin{pmatrix}
S_x \\
S_y \\
S_z
\end{pmatrix}
\end{eqnarray}
In the rotated coordinate the alignment of the spin becomes \(\langle\mathbf{S}_l\rangle = (S_0, 0,0)^\prime\). 
The rotation matrix \(\mathcal{R}(\theta, \phi)\) is the product of rotations around the \(z\)-axis by angle \(\phi\) and the \(y\)-axis by angle \(\theta\):
\begin{eqnarray}
\mathcal{R}(\theta, \phi) =
\begin{pmatrix}
\cos \theta \cos \phi & -\sin \phi & \sin \theta \cos \phi \\
\cos \theta \sin \phi & \cos \phi & \sin \theta \sin \phi \\
-\sin \theta & 0 & \cos \theta
\end{pmatrix}
\end{eqnarray}

To analyze small deviations from the perfectly aligned spin configuration (the ground state), the Holstein-Primakoff transformation\cite{HP1,HP2,HP3} is applied. This transformation introduces bosonic operators to represent spin deviations (magnons). The spin operators in terms of bosonic creation (\(b_l^\dagger\)) and annihilation (\(b_l\)) operators are:
\begin{eqnarray}
\begin{rcases}
&S_l^+ = S_l^{y^\prime} + i S_l^{z^\prime} = \sqrt{2S_0} \left(1 - \frac{b_l^\dagger b_l}{2S}\right)^{1/2} b_l
\\
&S_l^- = S_l^{y^\prime} - i S_l^{z^\prime}  = b_l^\dagger \sqrt{2S_0}  \left(1 - \frac{b_l^\dagger b_l}{2S}\right)^{1/2}
\\
&S_l^{x^\prime} = S_0 - b_l^\dagger b_l
\end{rcases}.
\end{eqnarray}
Here, \(S_l^+\) and \(S_l^-\) are the spin raising and lowering operators, respectively, which manipulate the spin state by adding or removing a magnon. \(S_l^{x^\prime}\) gives the longitudinal component of the spin, reduced by the number of magnons.
The Hamiltonian for FI can now be written as:
\begin{eqnarray}
\mathcal{H}_{\text{FI}} = -\mathcal{J} \sum_{\langle l, m \rangle} \mathbf{S}_l \cdot \mathbf{S}_m - g \mu_B h_{\text{dc}} \sum_l S_l^{x^\prime}
\end{eqnarray}
The first term represents the exchange interaction between neighboring spins, and the second term is the Zeeman energy due to an external magnetic field whose \(x^\prime\) -component is \({h}_{\text{dc}}\).
In momentum space, the Hamiltonian is diagonalized using the Fourier transform of the bosonic operators:
\[
{b}_{\bf q} = \frac{1}{\sqrt{N}} \sum_l e^{-i \mathbf{q} \cdot \mathbf{r}_l} b_l,\quad {b}_{\bf q}^\dagger = \frac{1}{\sqrt{N}} \sum_j e^{i \mathbf{q} \cdot \mathbf{r}_l} b_l^\dagger,
\]
leading to the diagonalized form:
\begin{eqnarray}
\mathcal{H}_{\text{FI}} = \sum_\textbf{q} \hbar \omega_\textbf{q} b_\textbf{q}^\dagger b_\textbf{q}
\end{eqnarray}
where \(\hbar \omega_\textbf{q}\) represents the energy of the magnon with wave vector \(\textbf{q}\).

The dynamic properties of the spin in FI are studied using the retarded Green's function, which provides information about the magnon propagator:
\begin{eqnarray}G(\mathbf{k}, t) = \frac{1}{i \hbar} \theta(t) \left[ S_\textbf{k}^+(t), S_\textbf{k}^-(0) \right]
\end{eqnarray}

At \(\textbf{q}=0\) and in the frequency domain, the Green's function simplifies to:
\begin{eqnarray}
G(\mathbf{q} = 0, \omega) \approx \frac{2S_0 / \hbar}{\omega - \omega_{\mathbf{q}=0} + i\,[\alpha_G + \delta\alpha_G(\omega,T)]\,\omega}
\end{eqnarray}
where \(\omega_\textbf{q} = D\vert\textbf{q}\vert^2 + g\mu_B h_{\text{dc}}\) is the magnon dispersion relation, with \(D\) being the spin-wave stiffness.

\subsection{Interfacial Exchange}
The interaction at the interface between 2DHG and FI is mediated by the exchange interaction. This interaction plays a crucial role in determining the spin dynamics and the overall magnetic properties of the system. The interfacial exchange Hamiltonian\cite{2DEG} can be described by the coupling between the spins of the 2DHG and the spins of the FI.
\begin{equation}
\mathcal{H}_{\text{ex}} = \sum_{\boldsymbol{q,k}} \mathcal{T}_{\boldsymbol{q},\boldsymbol{k}} \hat{s}_\mathbf{k}^+ \hat{S}_\mathbf{q}^- + \rm{H.c.}
\end{equation}
This Hamiltonian describes the interaction between the spin operators of the 2DHG (\( \hat{s}_{\bf k}^+ \)) and the FI (\( \hat{S}_\mathbf{q}^- \)), with \( \mathcal{T}_{\mathbf{q},\mathbf{k}} \) being the matrix element that quantifies the strength of this interaction for given momentum states \( \textbf{q} \) and \( \mathbf{k} \). The Hermitian conjugate (H.c.) term ensures that the Hamiltonian is Hermitian, accounting for the reverse processes.
The transition probability \( |\mathcal{T}_{\mathbf{q},{\bf k}}|^2 \) represents the probability of a spin-flip process at the interface, and it can be decomposed into two components:
\begin{equation}
|\mathcal{T}_{\mathbf{q},{\bf k}}|^2 = \mathcal {T}_1^2 \delta_{\mathbf{q},0} + \mathcal{T}_2^2
\end{equation}
Here, \( \mathcal{T}_1 \) corresponds to the ``clean'' processes where momentum is conserved (\( \delta_{\mathbf{q},0} \) ensures this conservation), and \( \mathcal{T}_2 \) accounts for the ``dirty'' processes that include momentum relaxation due to impurities or other scattering mechanisms at the interface.

In a similar way, the spin components \( (s_x, s_y, s_z) \) in the rotated frame \( (s_{x^\prime}, s_{y^\prime}, s_{z^\prime}) \) is written as:
\begin{equation}
\begin{pmatrix}
s_{x^\prime}\\
s_{y^\prime} \\
s_{z^\prime}
\end{pmatrix}
=
\mathcal{R}(\theta, \phi)
\begin{pmatrix}
s_x \\
s_y \\
s_z
\end{pmatrix}
\end{equation}
The spin operator \( {\textbf{s}}^a_\textbf{k} \) (\(a\equiv(x,y,z)\)) for the 2DHG in momentum space is expressed in terms of the creation (\( c_{\textbf{q},\sigma}^\dagger \)) and annihilation (\( c_{\mathbf{k} + \mathbf{q},\sigma} \)) operators for spin states \( \sigma \):
\begin{equation}
{\textbf{s}}^a_\mathbf{k} = \sum_{\sigma \sigma^\prime} \sum_\textbf{q} c_{\textbf{q},\sigma}^\dagger ({\boldsymbol\sigma_a})_{\sigma \sigma^\prime} c_{\textbf{k} + \textbf{q},\sigma}.
\end{equation}
The Pauli matrices \( {\boldsymbol\sigma_a} \) mediate the spin interaction between different spin states.
Finally, the spin-flip operators in the rotated frame are defined as:
\begin{equation}
\begin{rcases}
s_{x^\prime}^{\textbf{k},+}= s_{y^\prime}^{\textbf{k}} + i s_{z^\prime}^{\textbf{k}}\\
s_{x^\prime}^{\textbf{k},-}=({s_{x^\prime}^{\textbf{k},+}})^\dagger
\end{rcases}.
\end{equation}
These operators describe the raising (\( s_{x^\prime}^{\textbf{q},+} \)) and lowering (\( s_{x^\prime}^{\textbf{q},-} \)) of spin states in the rotated coordinate system, crucial for describing the dynamics of spin-flip processes at the interface.
\section{Calculation of the Gilbert damping factor}\label{sec2}
We adopt the Green's function formalism \cite{Gen tatara} to model the spin pumping effect, providing a rigorous framework for describing spin dynamics \cite{spin dynamics1, spin dynamics2} and the resulting enhancement of the Gilbert damping constant. In this section, we calculate the correction to the Gilbert damping factor due to interfacial exchange interactions in 2DHG coupled to FI, as illustrated in Fig.~\ref{Fig1}. The interfacial exchange modifies the electronic structure by introducing an impurity potential (refer to Appendix. [\ref{AppI}]), leading to self-energy corrections $\Sigma(\textbf{k},i\omega_m)$ that account for spin-flip scattering and renormalization of quasiparticle states. The presence of spin-orbit coupling (SOC) in the 2DHG further enhances spin-dependent scattering, significantly influencing magnetization dynamics.

To analyze these effects, we start with the Green’s function in the presence of second-order perturbation \cite{Flensberg,Coleman}, given by:
\begin{equation}
\mathcal{G}(\mathbf{k}, i\omega_m) = \frac{1}{(\mathcal{G}_0(\mathbf{k}, i\omega_m))^{-1} - \Sigma(\mathbf{k}, i\omega_m)},
\end{equation}
where $\mathcal{G}_0(\mathbf{k}, i\omega_m)$ is the bare Green's function, and $\Sigma(\mathbf{k}, i\omega_m)$ represents the self-energy. The imaginary part of $\Sigma(\mathbf{k}, i\omega_m)$ determines the broadening of energy levels due to spin fluctuations, directly affecting the dissipation of magnetization dynamics. This correction leads to the enhancement of the Gilbert damping factor, $\delta\alpha_G(\omega, T)$, which is derived from $\text{Im} \, \Sigma(\mathbf{q} = 0, \omega)$.

The self-energy \cite{2DEG}, incorporating spin-spin correlations and spin-flip scattering, is given by:
\begin{eqnarray}
\Sigma(\mathbf{k}, i\omega_n) = \frac{|\mathcal{T}_1|^2}{4\beta} &\sum_{\mathbf{k}^\prime, i\omega_m} \text{Tr} [{\sigma}_{x^\prime}^- \mathcal{G}(\mathbf{k}^\prime, i\omega_m) \nonumber\\
&\times{\sigma}_{x^\prime}^+ \mathcal{G}(\mathbf{k}^\prime + \mathbf{k}, i\omega_m + i\omega_n) ].
\end{eqnarray}
The Green's function in momentum space for the Matsubara frequency $i\omega_m$ takes the form:
\begin{eqnarray}
\mathcal{G}({\bf k}, i\omega_m) = \frac{\mathcal{A}({\bf k},i\omega_m) \mathbf{\sigma_0} - \mathbf{h}_{\text{SOC}} \cdot \boldsymbol{\sigma}}{\mathcal{B}({\bf k},i\omega_m)},
\end{eqnarray}
where the functions $\mathcal{A}({\bf k},i\omega_m)$ and $\mathcal{B}({\bf k},i\omega_m)$ are defined as:
\begin{eqnarray}
\mathcal{A}({\bf k},i\omega_m) = i\hbar\omega_m - \zeta(k) + i\frac{\Gamma \, \text{sgn}(\omega_m)}{2},
\end{eqnarray}
\begin{eqnarray}
\mathcal{B}({\bf k},i\omega_m) = \prod_{s = \pm} \left(i\hbar\omega_m - E_s(\textbf{k}) + i\frac{\Gamma \, \text{sgn}(\omega_m)}{2}\right).
\end{eqnarray}\label{Gamma}
\begin{figure}[b]\includegraphics[width=70.5mm,height=75.5mm]{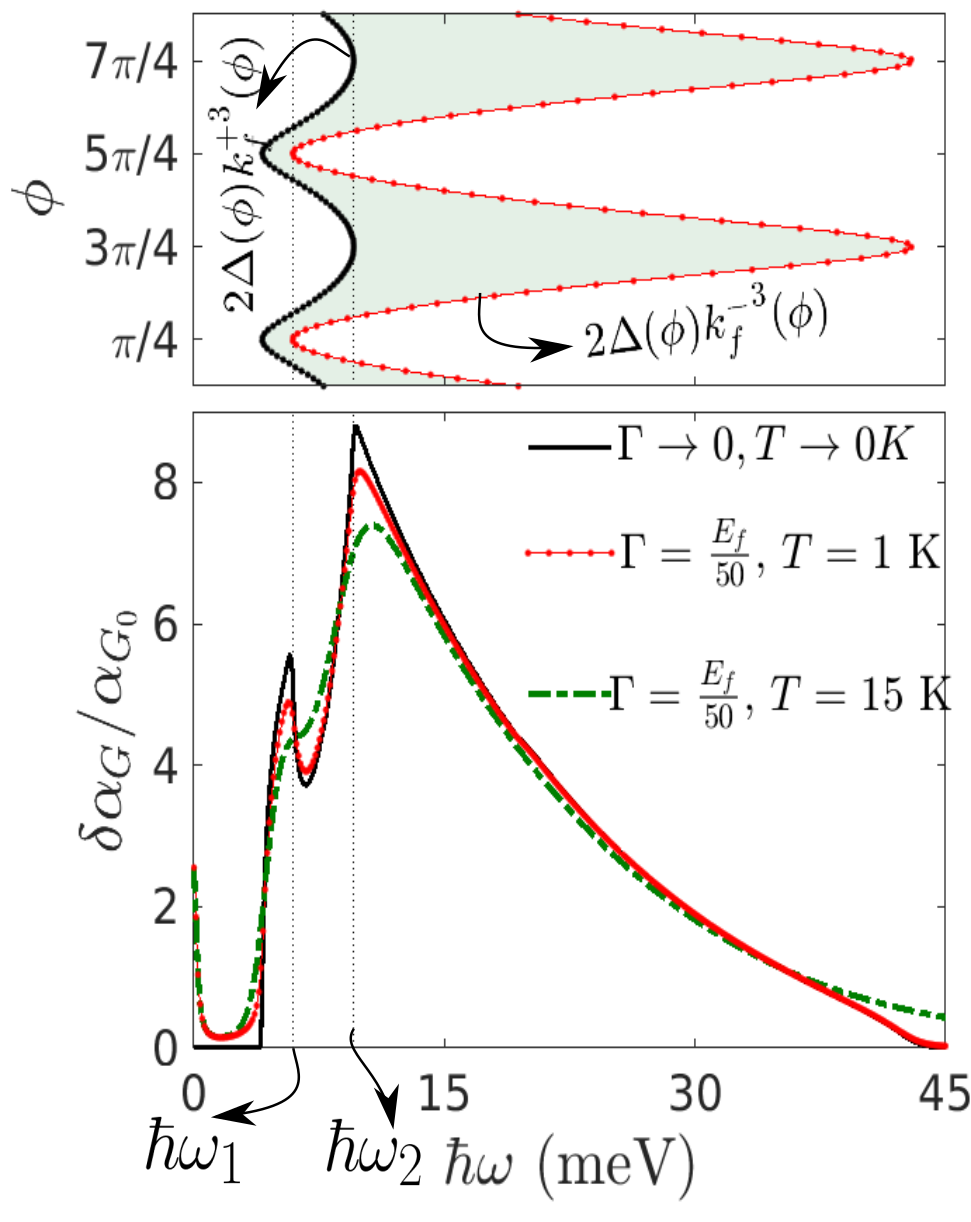}	
	\caption{\textbf{Upper pannel}: Variation of the quantity $2\vert {\bf h}_{\rm SOC}({\bf k})\vert_{(k_f^\pm(\phi),\phi)}$ appearing in the delta function at Eq. [\ref{Gilbert}] as a function of azimuthal angle $\phi$. \textbf{Lower pannel}: Frequency dependent Gilbert damping parameter with magnon broadening effect, highlighting the impact of temperature on damping behaviour for  $\alpha/\beta=2$, $E_f=20$ meV. The critical frequencies $\hbar\omega_1$ and $\hbar\omega_2$ corresponds to van-Hove singularities where enhanced optical transitions occur, significantly influencing the damping response.} 
	\label{Fig3}
\end{figure}
By explicitly evaluating the self-energy, we obtain:
\begin{widetext}
\begin{eqnarray}\label{Sel_q0}
&&\Sigma(\textbf{q} = 0, i\omega_m) = \frac{|\mathcal{T}_1|^2}{\beta}\sum_{k, i\omega_n} \frac{\mathcal{A} ({\bf k},i\omega_m) - \hat{\textbf{h}}_{\text{SOC}} \cdot {\textbf{m}}}{\mathcal{B}({\bf k},i\omega_m) } \nonumber \frac{\mathcal{A}({\bf k+k^\prime},i(\omega_m+\omega_n)) + \hat{\textbf{h}}_{\text{SOC}} \cdot {\textbf{m}}}{\mathcal{B}({\bf k+k^\prime},i(\omega_m+\omega_n))}.
\end{eqnarray}
\end{widetext}
A detailed derivation of this expression is provided in Appendix [\ref{AppII}].
The enhancement in the Gilbert damping coefficient \cite{Kato3}, denoted as $\delta\alpha_G(\omega,T)$, is now determined using $\delta \alpha_G(\omega,T) \approx -\frac{2S_0}{\hbar \omega} \operatorname{Im} \, \Sigma \,(\textbf{q} = 0,i\omega_m)$ as following
\begin{widetext}
\begin{eqnarray}\label{Gilbert}
\frac{\delta\alpha_{G}(\omega,T)}{\alpha_{G,0}}\ &=&  -\sum_{s,s^\prime=\pm1}\int_{0}^{\infty} d\zeta \int_0^{2\pi} \frac{d\phi}{2\pi} \, \delta_L[\hbar\omega + (s - s^\prime) \vert{\bf h}_{\rm{SOC}}({\bf k})\vert]\nonumber \,\frac{1 - s\, \hat{\textbf{h}}_{\text{SOC}}({\bf k}) \cdot {\textbf{m}}}{2} \frac{1 + s^\prime\,\hat{\textbf{h}}_{\text{SOC}}({\bf k}) \cdot {\textbf{m}}}{2} 
\nonumber\\ 
&\times&\left[  \left( f(E_-(\textbf{k}))\delta_{ss^\prime} \right) 
+  \left( f(E_{s}(\textbf{k})) - f(E_{s^\prime}(\textbf{k})) \right)(1 - \delta_{ss^\prime}) \right].
\end{eqnarray}
\end{widetext}
where the dimensionless parameter $
\alpha_{G,0} =2\pi S_0\, |\mathcal{T}_1|^2\, \Omega\,\mathcal{D}(E_f),
$
and $\delta_L(x)$ represents the Lorentzian delta function,
\(
\delta_L(x) = {\Gamma/2}/(\pi{x^2 + (\Gamma/2)^2}).
\) The detailed derivation of the above Eq. is provided in Appendix[\ref{AppIII}]. From the form of \(\alpha_{G,0}\), it is evident that the density of states plays a crucial role in determining the efficiency of the enhanced Gilbert damping strength. In a previous study \cite{2DEG}, spin pumping was explored in 2DEG system while our study investigates a heterostructure with 2DHG. By comparing their density of states, we conclude that the damping coefficient strength in the 2DHG case is order larger than in 2DHG system. Now, using Eq. [\ref{Gilbert}], we proceed with numerical evaluations to analyze the impact of interfacial exchange on the Gilbert damping factor. These results provide quantitative insights into the role of spin-flip scattering and spin-orbit coupling in enhancing damping in the 2DHG system.
\begin{figure}[http!]		\includegraphics[width=85.5mm,height=40.5mm]{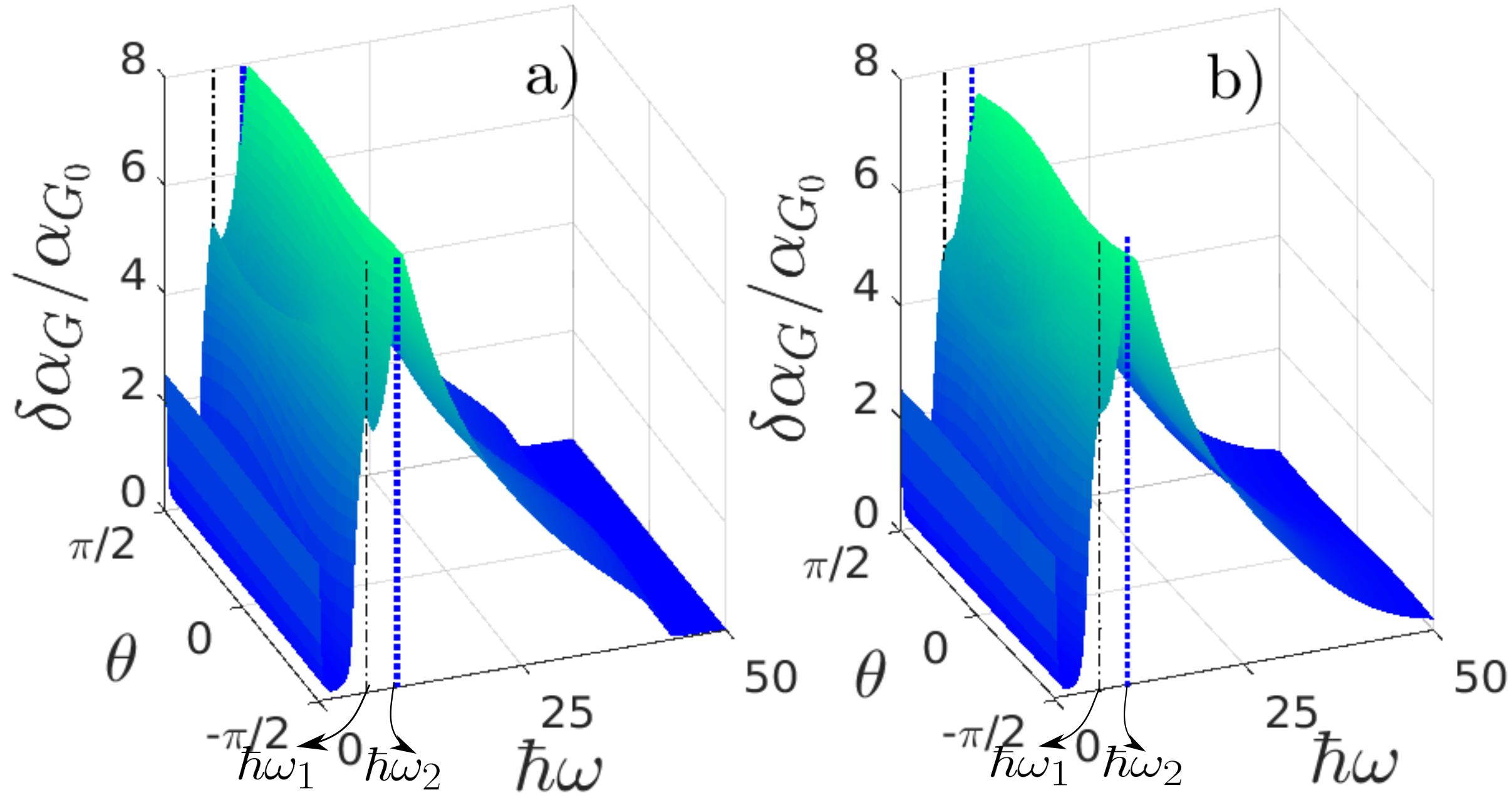}	
	\caption{Contour plot of the enhanced Gilbert damping in a 2DHG for  at two different temperatures ${\alpha}/{\beta}=2$ (a) T$\rightarrow$0 K (b) T=10 K, showing the shift in the peak at the characteristic frequency with the variation of temperature }
	\label{Fig4}
\end{figure}
\section{Results and discussion}\label{sec3}
\begin{figure}[http!]		\includegraphics[width=70.5mm,height=47.5mm]{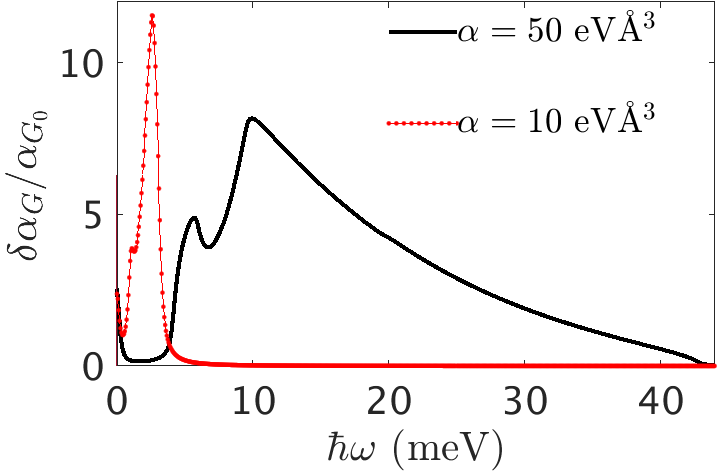}	
	\caption{The Gilbert damping enhancement $\delta\alpha_{G}$/$\alpha_{G,0}$ exhibits significant broadening as RSOC strength$\alpha$ increases. For $\alpha = 50 \, \text{eV} \mathring{\rm{A}}^3$ (black solid line), the damping spectrum extends over a wider frequency range compare to $\alpha = 10 \, \text{eV} \mathring{\rm{A}}^3$ (red dashed line) indicating that stronger RSOC leads to increased spectral broadening of Gilbert damping. }
	\label{Fig5}
\end{figure}


\noindent\textbf{1. Transition between \(s = -1\) and \(s^\prime = 1\):}  
At higher frequencies, the transition between different energy bands (inter-band) becomes more significant as the photon energy is large enough to excite spin across bands. The relevant transitions contributing to the optical response are between \( s = -1 \) and \( s^\prime = 1 \), as these correspond to the change in spin orientation (flip) during the interaction with the electromagnetic field. These transitions are governed by the energy difference \( \hbar\omega + (s - s^\prime) \vert \textbf{h}_{\rm{SOC}}\vert \). This transition  ensures that the spin-flip dynamics are properly accounted for, influencing the optical response.

\noindent\textbf{2. Intra-band vs. Inter-band Transitions:}  
The behavior of intra-band and inter-band transitions is determined by the spin-orbit coupling (SOC) and the energy dispersion. The term \( \hat{\textbf{h}}_{\text{SOC}}({\bf k}) \cdot{\textbf{m}} \) modulates the strength of these transitions. At low frequencies, the intra-band transitions are stronger because the hole energy required for transitions within the same band is small. This is reflected in the smaller energy scale \( \hbar\omega \) involved in the transition. 

\noindent\textbf{3.  Optical Transition due to Spin pumping:} 
In our analysis optical transition occurs in microwave frequency range $\hbar\omega_2\leq\hbar\omega\leq\hbar\omega_1$ where $\hbar\omega_1=2 \Delta(\phi)[k_f^{-}(\phi)]^3$ and $\hbar\omega_2=2 \Delta(\phi)[k_f^{+}(\phi)]^3$. Here $k_f^{-}(\phi)$ and $k_f^{+}(\phi)$ are Fermi wave vectors for different spin subbands. We have observed that within this range a significant enhancement in Gilbert damping coefficient occurs due to the presence of van-Hove singularities as shown in Fig-[\ref{Fig3}]. This singularity manifests the conductivity-like optical transition, particularly when the broadening parameter $\Gamma\rightarrow0$, results in a peak exactly at one of the characteristic transition frequencies. The specific transition frequencies where the singularities occur are given by the following condition.  
\begin{eqnarray*}
	\hbar\omega =
	\begin{cases} 
		\hbar\omega_1, & \text{if } \phi = \frac{\pi}{4}, \frac{5\pi}{4} \\ 
		\hbar\omega_2, & \text{if } \phi = \frac{3\pi}{4}, \frac{7\pi}{4} 
	\end{cases}
\end{eqnarray*}
Physically this means that at these specific microwave frequencies and azimuthal angular positions, the number of available states for transition is maximized (Refer to the upper panel in Fig. [\ref{Fig3}]), leading to enhanced Gilbert damping parameter. In the context of spin pumping, it directly impacts the spin current generation by modifying the absorption characteristics which depends on the RSOC strength $\alpha$. The presence of these singularities indicates a critical point in the density of states where the transition probability is significantly enhanced. The inclusion of finite broadening smooths out the singularity but still preserves a peak structure highlighting the strong dependence of spin transport on magnon interaction and its lifetime ($\Gamma\propto1/\tau$). The magnon lifetime $\tau$ defines the duration over which a magnon remains coherent before decaying due to interaction.  A shorter magnon lifetime leads to increased spectral broadening. In this case, the broadening is influenced by the cubic SOC and the interfacial exchange interaction which impacts the enhancement in Gilbert damping. In Fig-[\ref{Fig4}], increased Gilbert damping parameter $\delta\alpha_G$(black curve) is shown with infinitesimally small broadening which implies weaker magnon interaction with longer lifetime at negligibly small temperature. 

\subsection{Temperature effect on Gilbert damping}
In Fig-[\ref{Fig4}], we observed that with the increment of the temperature leads to the decrement in enhanced Gilbert damping coefficient. However, its magnitude remains larger compared to the conventional two-dimensional electron gas(2DEG), which can be attributed to the presence of cubic RSOC. At low temperatures, $\delta\alpha_G$ is strongly determined by the spin momentum locking induced by cubic RSOC and the weak influence of the interaction with magnon zero mode. As temperature increases, thermal excitations introduce more thermally activated magnon scattering events which are captured by the second-order perturbation theory discussed in Appendix-[\ref{AppI}]. Since the magnon population starts to increase, the stronger magnon interaction is leading to a shift in the characteristic frequency($\hbar\omega_1$, $\hbar\omega_2$) at which the peak in enhanced Gilbert damping appears. This shift suggests that the dominant energy scales(refer to Eq.[\ref{Gilbert}]) governing spin dissipation change as thermal effects which modify the magnon density and the available states for spin scattering processes. These processes disrupt the coherent precession, thereby reducing the Gilbert damping enhancement compared to its value at $T\rightarrow 0$ K. 

\subsection{Role of the Fermi level}
Additionally, the spin Fermi level $E_f$ governs the spin current by introducing a spin imbalance between spin-up and spin-down bands. A non-zero $E_f$ ensures a finite spin current, as setting $E_f= 0$ eliminates the spin imbalance, thereby quenching the spin current. If $E_f= 0$, it signifies the absence of spin imbalance\cite{Kato2} between spin-up and spin-down holes, resulting in no net spin accumulation and no spin current injection into the non-magnetic material (NM). Without spin accumulation, the driving force for spin transport from the FI vanishes, leading to the disappearance of the Gilbert damping enhancement, which originates from angular momentum loss via the spin current. This behavior starkly differentiates the system from a conventional 2DEG, where spin imbalance and associated spin currents manifest differently. In this system, the intricate dependence on spin accumulation highlights its distinct dynamics, particularly under the influence of temperature and RSOC.  
\subsection{Impact of RSOC on spectral broadening and Gilbert damping enhancement}
We observed in Fig-[\ref{Fig5}] that stronger RSOC influences the spectral broadening of Gilbert damping. This occurs because the RSOC modifies the dispersion of the magnon by expanding the phase space for magnon interaction and thereby reducing the magnon's lifetime. As the RSOC increases, the broadening of the spectral function also increases allowing Gilbert damping to extend over a wider range of frequency. Consequently, the spin angular momentum dissipation becomes more significant over a wider range of frequencies. However, the enhancement in Gilbert damping weakens as RSOC becomes more pronounced.

\section{Conclusions and Summary}\label{sec4}
In summary, our study highlights the crucial role of cubic RSOC in enhancing Gilbert damping at the 2DHG/FI interface, where the 2DHG acts as the spin-absorbing layer for spin pumping from the FI. The presence of cubic RSOC facilitates efficient spin-orbit scattering, increasing spin-pumping efficiency compared to conventional 2DEGs and amplifying angular momentum transfer. Our results demonstrate that although $\delta \alpha_G$ decreases with temperature, it remains larger than in 2DEGs due to the persistent influence of cubic RSOC. The interplay between RSOC and magnon absorption leads to spectral broadening, modifying spin relaxation dynamics and shifting the characteristic frequency of the damping peak. This shift reflects a thermally driven redistribution of spectral weight due to magnon absorption, renormalizing spin dissipation processes.
The primary contribution to Gilbert damping arises from interband transitions, expected to exhibit conductivity-like behavior in the limits $T \to 0$ and $\Gamma \to 0$. Beyond temperature effects, we find that RSOC strength, tunable via an external electric field, directly influences the magnitude of Gilbert damping. A stronger gate field enhances RSOC, promoting increased spin relaxation and more efficient spin momentum transfer, whereas a weaker field suppresses SOC, reducing damping. This electric field tunability provides a pathway for real-time control over spin current flow, enabling voltage-controlled spintronic devices where the electric field acts as a ``spin current regulator''.
Furthermore, our findings emphasize the necessity of a finite Fermi level ($E_f$) in maintaining spin imbalance, which is essential for spin transport and dissipation. Setting $E_f = 0$ eliminates spin imbalance and suppresses damping, underscoring the fundamental link between spin pumping, temperature effects, and the Fermi level. These insights pave the way for tuning spin dynamics via electric and thermal means, advancing spintronic technologies based on cubic RSOC-engineered heterostructures.

	\textit{Acknowledgments}: This work is an outcome of the Research work carried out under the SRG Project, SRG/2023/001516, 
	Anusandhan National Research Foundation (ANRB),
	Government of India
	
	\appendix
	 \section{ Impurity scattering }\label{AppI}
	 In our study, we have considered the impurity potential as
	\begin{eqnarray}
	\mathcal{ V}_{\rm imp}(\textbf{r}) = \sum_{i=1}^{N_{\text{imp}}} u(\textbf{r} - \mathbf{R}_i),
	\end{eqnarray}
	
\noindent	where $\textbf{R}_i$ is the position vector of the randomly distributed impurities. Now  we assume $\frac{n_{\text{imp}}}{n_{\text{hole}}} \ll 1$ which indicating the presence of weak scattering potential. This scattering potential \( u(r - \mathbf{R}_i) \) deviates significantly from zero only when $|r - \mathbf{R}_i| < a $ where $a$ is the characteristic length scale where the scattered particle effectively interacted with the potentianl and beyond that range, the potential diminishes rapidly. To incorporate the self-energy correction  $\gamma(\mathbf{k}, \omega)$ into the Green's function due to this potential, we have utilized the Dyson equation\cite{Flensberg}
	 	 \begin{figure}[hhtp!]		\includegraphics[width=60.5mm,height=15.5mm]{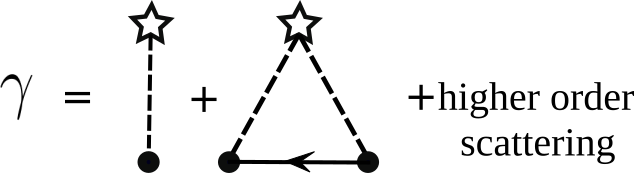}	
	 	 	\caption{Feynman diagram illustrating the perturbations with respect to the impurity potential and its corresponding diagrammatic expansion.}
	 	 	\label{Fig6}
	 	 \end{figure}
	 \begin{eqnarray*}
	 	\mathcal{G}_{\text{imp}}(\mathbf{k}, \omega) = \mathcal{G}_0(\mathbf{k}, \omega) + \mathcal{G}_0(\mathbf{k}, \omega) \gamma(\mathbf{k}, \omega) \mathcal{G}_{\text{imp}}(\mathbf{k}, \omega).
	 \end{eqnarray*}

\noindent	 In Fig[\ref{Fig6}] the irreducible Feynman diagrams up to the second order are considered. Here using the Born approximation, the self-energy\cite{Flensberg,Coleman} was then calculated as 	
	 \begin{eqnarray*}
	 \gamma(\textbf{k},i\omega_m)=\sum_\textbf{k}\langle \textbf{k}|\mathcal{V}_{\text{imp}}|\textbf{k}^\prime\rangle\mathcal{G}_0(\textbf{k},\omega)\langle \textbf{k}^\prime|\mathcal{V}_{\text{imp}}|\textbf{k}\rangle
	 \end{eqnarray*}
 here $\mathcal{G}_0(\textbf{k},\omega)$ describes the propagation in between scattering events. Thus we calculate the self energy due to the impurity scattering as follows 
 \begin{eqnarray*}
 \gamma\left(\textbf{k}, i\omega_m \right) = n_{\text{imp}} u^2\int \frac{d^2k}{(2\pi)^2} \mathcal{G}_0(\textbf{k}, i\omega_m),
 \end{eqnarray*}
 
	 The results are obtained in the Born approximation, where the self-energy is calculated as 
	 \begin{eqnarray*}
	 	\gamma\left(\textbf{k}, i\omega_m \right) = -i \frac{n_{\text{imp}}u^2 k_f}{2 v_f} \, \text{sgn}(\omega_m) \mathbf{\sigma_0} \equiv -i \frac{\Gamma}{2} \, \text{sgn}(\omega_m)\mathbf{\sigma_0} ,
	 \end{eqnarray*}
 where $\Gamma = \frac{n_{\text{imp}} u^2 k_f}{v_f}$ is the scattering rate, representing the impurity strength.
	
	 \section{For deriving Eq. [\ref{Sel_q0}]}\label{AppII}
Here, we derive {Eq. (\ref{Sel_q0})} by systematically rewriting and simplifying the relevant expressions. We begin by expressing the Green’s function of the conduction holes in the following form\cite{2DEG}:  
\begin{eqnarray}
\mathcal{G}(\textbf{k}, i\omega_m) = \frac{1}{\mathcal{B}(i\omega_m)} \left[\mathcal{A}(i\omega_m) \sigma_0 + \textbf{h}_{\rm SOC} \cdot \boldsymbol{\sigma} \right].
\end{eqnarray}
 Here \(\mathcal{A}(i\omega_m)\equiv \mathcal{A}(\textbf{k},i\omega_m)\) and \(\mathcal{B}(i\omega_m)\equiv \mathcal{B}(\textbf{k},i\omega_m)\).
Next, we rewrite the trace term $I$ in as:  
\begin{eqnarray}
I \equiv \text{Tr}\left[{\sigma}_x^- \, {\mathcal{G}}(\textbf{k}, i\omega_m) \, {\sigma}_x^+ \, {\mathcal{G}}(\textbf{k}, i\omega_m + i\omega_n)\right].
\end{eqnarray}
Using the explicit form of the Green’s function ${\mathcal{G}}(\textbf{k}, i\omega_m)$, the trace can be expressed as:  
\begin{eqnarray}
I &=& \frac{1}{\mathcal{B}(i\omega_m)\mathcal{B}(i(\omega_m+\omega_n))} \, \text{Tr}\big[(\mathbf{a}^* \cdot \boldsymbol{\sigma})\big(\mathcal{A}(i\omega_n) \mathbf{\sigma_0} + \mathbf{b} \cdot \boldsymbol{\sigma}\big) \nonumber\\
&\times& (\mathbf{a} \cdot \boldsymbol{\sigma})\big(\mathcal{A}(i(\omega_m+\omega_n)) \mathbf{\sigma_0} + \mathbf{b} \cdot \boldsymbol{\sigma}\big)\big].
\end{eqnarray}
Specifically, we define:
\begin{equation}
\mathbf{a} = \left(-\sin \theta, \cos \theta, i\right), \quad \textbf{b} = -\textbf{h}_{\rm SOC}.
\end{equation}
To simplify this, we employ the Pauli matrix identity:  
\begin{equation}
(\mathbf{a} \cdot \boldsymbol{\sigma})(\mathbf{b} \cdot \boldsymbol{\sigma}) = (\mathbf{a} \cdot \mathbf{b}) \mathbf{\sigma_0} + i(\mathbf{a} \times \mathbf{b}) \cdot \boldsymbol{\sigma},
\end{equation}
where $\mathbf{a} \cdot \mathbf{b}$ is the scalar product, $\mathbf{a} \times \mathbf{b}$ is the vector cross product, and $\mathbf{\sigma_0}$ represents the identity matrix in spin space.  
Substituting this identity into the trace and expanding the terms gives:
\begin{align}
I = \frac{2}{\mathcal{B}(i\omega_m)\mathcal{B}(i(\omega_m+\omega_n))} \big[ \mathcal{A}(i\omega_m) \mathcal{A}(i(\omega_m+\omega_n)) (\mathbf{a}^* \cdot \mathbf{a}) \nonumber\\
+ i \mathcal{A}(i(\omega_m+\omega_n)) (\mathbf{a}^* \times \mathbf{b}) \cdot \mathbf{a} + i \mathcal{A}(i\omega_m) (\mathbf{a}^* \cdot (\mathbf{a} \times \mathbf{b})) \nonumber\\
- (\mathbf{a}^* \times \mathbf{b}) \cdot (\mathbf{a} \times \mathbf{b}) + (\mathbf{a}^* \cdot \mathbf{b})(\mathbf{a} \cdot \mathbf{b}) \big].
\end{align}
Finally, substituting the explicit forms of $a$ and $b$, we obtain {Eq. (\ref{Sel_q0})}. 

	 	 \section{FROM MATSUBARA TO RETARDED GREEN'S FUNCTION:AN ANALYTICAL CONTINUATION APPROACH}\label{AppIII}
We use the analytic continuation in order to evaluate the Matsubara summation\cite{Flensberg, Coleman} in Eq. [\ref{Sel_q0}]. We first write the following ratio
	 \begin{eqnarray}
\frac{\mathcal{A}}{\mathcal{B}} = \frac{1}{2} \sum_{s = \pm} \frac{1}{i \hbar \omega_m - E_s(\textbf{k}) + i \frac{\Gamma}{2} \, \text{sgn}(\omega_m)},
\\	\frac{\hat{\bf h}_{\text{SOC}} \cdot {\textbf{m}}}{\mathcal{B}} = \frac{1}{2} \sum_{s = \pm} \frac{s \, \hat{\bf h}_{\text{SOC}} \cdot {\textbf{m}}}{i \hbar \omega_m - E_s(\textbf{k}) + i \frac{\Gamma}{2} \, \text{sgn}(\omega_m)}
\end{eqnarray}
\begin{figure}[b]
	\includegraphics[width=50.5mm,height=31mm]{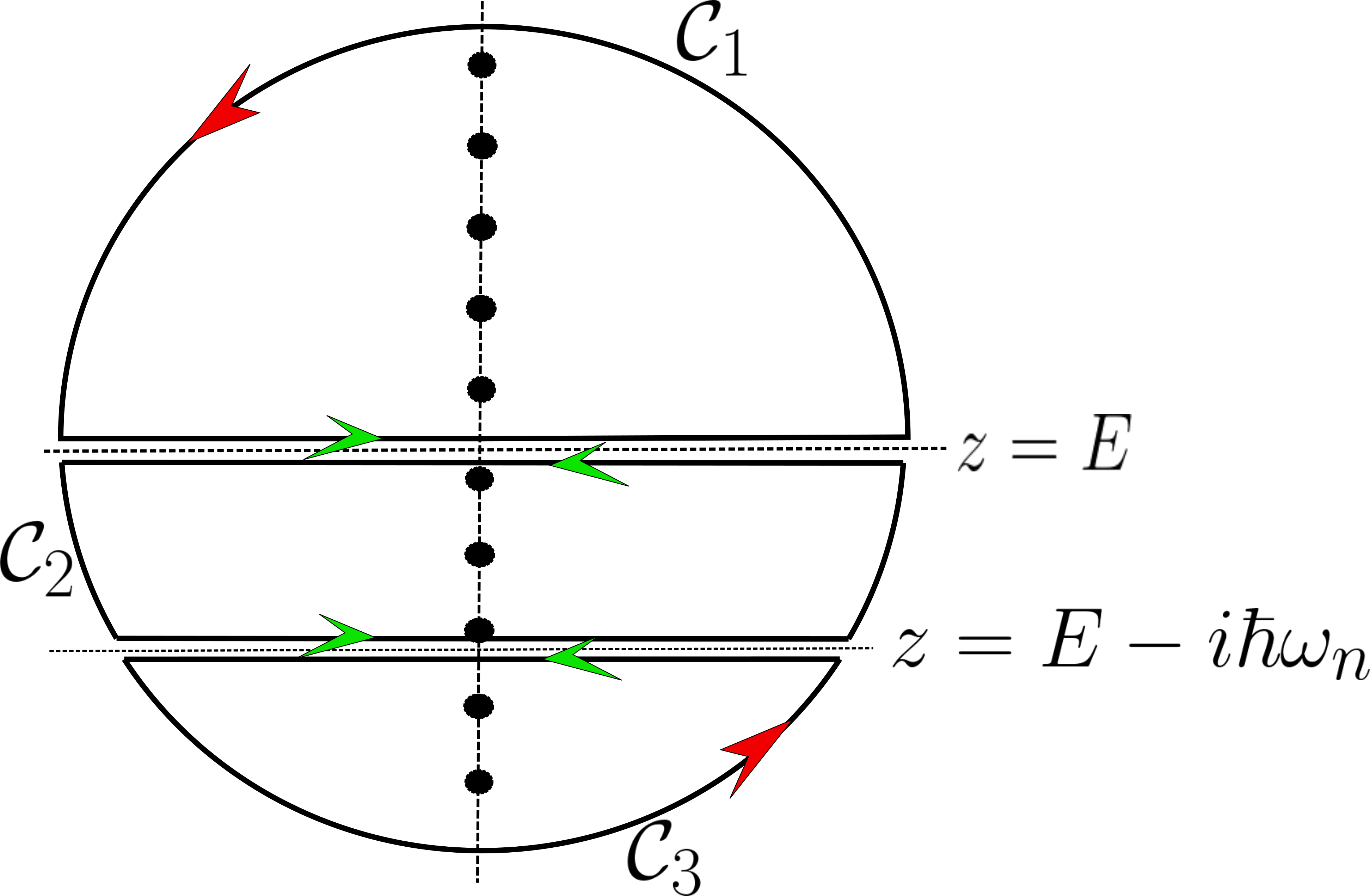}
	\caption{Contour representation for computing the Matsubara frequency summation in Eq. [\ref{A2}]. The integration contours \(\mathcal{C}_1\), \(\mathcal{C}_2\), and \(\mathcal{C}_3\) are set by the branch cuts at \({\rm Im}(z) = 0\) and \({\rm Im}(z) = -i\hbar\omega_n\). Poles along the imaginary axis correspond to the Fermi-Dirac distribution, aiding the analytical continuation. in Eq. [\ref{b8}].}
	\label{FigA1}
\end{figure}
\noindent A similar expression can also be written for the other corresponding term that appears in Eq. [\ref{Sel_q0}].
The self-energy of the composite system in the limit \({\bf q}\rightarrow 0\) can now be written as
\begin{eqnarray}\label{C3}
&\Sigma(\textbf{q}=0, i\omega_m) = \frac{|\mathcal{T}_1|^2}{4} \sum_{s,s^\prime=\pm,{\bf k}}  (1 - s \hat{\bf h}_{\text{SOC}} \cdot {\textbf{m}})\nonumber \\&\times (1 + s^\prime \hat{\bf h}_{\text{SOC}} \cdot {\textbf{m}})I^{ss^\prime}(\textbf{k}),
\end{eqnarray} 
where
\begin{eqnarray}\label{A2}
&&I^{ss^\prime}(\textbf{k})=\frac{1}{\beta}\sum_{m\in \mathbb{Z}}G_0^s({\bf k},\omega_m)G_0^{s^\prime}({\bf k},\omega_m+\omega_n)\nonumber\\&&=\frac{1}{\beta}\sum_{m\in \mathbb{Z}}\frac{1}{i\omega_m-E_s({\bf k})+i\frac{\Gamma}{2}{\rm sgn}(\omega_m)}\nonumber\\&&\times\frac{1}{i(\omega_m+\omega_n)-E^{s^\prime}({\bf k})+i\frac{\Gamma}{2}{\rm sgn}(\omega_m+\omega_n)}.
\end{eqnarray}

\noindent\textit{\textbf{{Analytical continuation}}}: To evaluate the summation over ($m\in \mathbb{Z}$) in Eq. [\ref{A2}],
we use the standard analytical continuation, $\frac{1}{\beta}\sum_{m=-\infty}^{\infty}g(i \omega_m)=\frac{1}{2\pi i}\oint_C dz f(z)g(z)$, with $f(z)=1/(e^{\beta z}+1)$, we have
\begin{eqnarray}\label{Iss}
&&I^{ss^\prime}(\textbf{k})=\frac{1}{2\pi i}\oint_C \frac{f(z)}{z-E_s({\bf k})+i\frac{\Gamma}{2}{\rm sgn}[{\rm Im}(z)]}\nonumber\\
&&\times\frac{1}{z+i\omega_n-E_{s^\prime}({\bf k})+i\frac{\Gamma}{2}{\rm sgn}[{\rm Im}(z)+\omega_n]}.
\end{eqnarray}
Consider the integral to be integrated over the closed contour as shown in Fig. [\ref{FigA1}]. 
In order to evaluate the integral on the right hand side of the above Eq. [\ref{Iss}], we modify the contour to be a sum of the closed paths $\mathcal{C}_1$, $\mathcal{C}_2$, and $\mathcal{C}_3$ in order to avoid the two mentioned branh-cuts (refer to Fig. [\ref{FigA1}]). By extending the contour to infinity, the integral in Eq. [\ref{Iss}] can be evaluated about the horizontal paths (shown by the green arrows). Thus we can change the integration variable in the above equation to $z =E + i\eta$ for the horizontal line in $\mathcal{C}_1$, ($z =E - i\eta$ and $z =E -i\omega_n+ i\eta$ about the two horizontal lines in  $\mathcal{C}_2$) and $z =E -i\omega_n- i\eta$ for the horizontal line in $\mathcal{C}_3$, where $\eta\rightarrow 0$ is a positive infinitesimal number. Thus the integral in Eq. [\ref{Iss}] is 
\begin{widetext}
	\begin{eqnarray}\label{b8}
	&&\oint_C \frac{f(z) dz}{(z-E_s(\textbf{k})+i\frac{\Gamma}{2}{\rm sgn}[{\rm Im}(z)])(z+i\omega_n-E_{s^\prime}(\textbf{k})+i\frac{\Gamma}{2}{\rm sgn}[{\rm Im}(z)+\omega_n])}=-\int_{-\infty}^\infty dE f(E)\nonumber\\&&\times
	\Big[\frac{i\Gamma}{(E-E_s(\textbf{k}))^2+(\Gamma/2)^2}\frac{1}{E+i\omega_n-E_{s^\prime}({\bf k})+i\Gamma/2}+\frac{i\Gamma}{(E-E_{s^\prime}(\textbf{k}))^2+(\Gamma/2)^2}\frac{1}{E-i\omega_n-E_s(\textbf{k})-i\Gamma/2}\Big].\nonumber\\
&&= -\frac{1}{\pi} \int dE \, f(E) \left[ 
\frac{1}{E + \hbar \omega - E_{s^\prime}(\textbf{k}) + i \Gamma / 2} \delta(E - E_s(\textbf{k})) 
+ \right. 
\left. \frac{1}{E - \hbar \omega - E_s(\textbf{k}) - i \Gamma / 2} \delta(E - E_{s^\prime}(\textbf{k})) 
\right]
\end{eqnarray}
Now the imaginary part of \(I^{ss^\prime}({\bf k})\) can be extracted as 
\begin{align*}
\mathrm{Im}[I^{ss^\prime}(\textbf{k})] = -\frac{1}{\pi} \left[(f(E_s(\textbf{k})) - f(E_{s^\prime}(\textbf{k})))\times\frac{-\Gamma/2}{(E_s(\textbf{k}) -E_{s^\prime}(\textbf{k}) + \hbar \omega)^2 + (\Gamma/2)^2}\right]
\end{align*}
\end{widetext}
Now we replace the summation over the wave vector(\textbf{k}) with an integral over momentum space which allows more convenient analytical tratment in the continuum limit,
\begin{eqnarray*}
\Omega \sum_\textbf{k} \approx\mathcal{D}(E_f) \int_{0}^{\infty} d\zeta \int_{0}^{2\pi} \frac{d\phi}{2\pi},
\end{eqnarray*}
such that we can rewrite Eq. [\ref{C3}] as 
\begin{widetext}
\begin{eqnarray}
&&\operatorname{Im}[\Sigma(\textbf{q}=0, i\omega_n)] = -\frac{2\pi S_0 \,\Omega |\mathcal{T}_1|^2}{4} 
\sum_{s,s^\prime = \pm}\mathcal{D}(E_f) 
\int_{0}^{\infty} d\zeta\nonumber\int_0^{2\pi} d\phi \, \delta_L\left[(s-s^\prime)\,\vert{\bf h}_{\rm{SOC}}({\bf k})\vert + \hbar \omega\right]
\nonumber\\ 
&\times&(1 - s \, \hat{\mathbf{h}}_{\text{SOC}} \cdot \hat{\mathbf{m}}) 
(1 + s^\prime \,\hat {\mathbf{h}}_{\text{SOC}} \cdot \hat{\mathbf{m}}) 
\left[  \left( f(E_-(\textbf{k}))\delta_{ss^\prime} \right) 
+  \left( f(E_{s}(\textbf{k})) - f(E_{s^\prime}(\textbf{k})) \right)(1 - \delta_{ss^\prime}) \right]
\end{eqnarray}
\end{widetext}




\end{document}